\documentclass[a4paper]{panl}

\usepackage{cite}
\usepackage{wrapfig}
\usepackage{graphicx}
\usepackage{amssymb}
\usepackage{amsfonts}
\usepackage{amsmath}
\usepackage{longtable}
\usepackage{rotating}
\usepackage{lscape}
\usepackage{epsfig}
\usepackage{multirow}

\originalTeX

\begin{document}

\issuearea{Physics of Elementary Particles and Atomic Nuclei. Theory}

\title{Model-independent femtoscopic L\'evy imaging for elastic  proton-proton scattering}

\maketitle

\authors{T.~Cs\"org\H{o}$^{\,\, a,b,c,}$ \footnote{E-mail: tcsorgo@cern.ch}, 
R.~Pasechnik$^{\,\, d,e,f,}$ \footnote{E-mail: Roman.Pasechnik@thep.lu.se},
A.~Ster$^{\,\, c,}$ \footnote{E-mail: Ster.Andras@wigner.mta.hu}
}

\from{$^{a}$\,CERN, CH-1211 Geneva 23, Switzerland}
\vspace{-3mm}
\from{$^{b}$\,EKE KRC, H-3200 Gy\"ongy\"os, M\'atrai \'ut 36, Hungary}
\vspace{-3mm}
\from{$^{c}$\,Wigner FK, H-1525 Budapest 114, P.O.Box 49, Hungary}
\vspace{-3mm}
\from{$^{d}$\,Department of Astronomy and Theoretical Physics, 
Lund University,\\ SE-223 62 Lund, Sweden}
\vspace{-3mm}
\from{$^{e}$\,Nuclear Physics Institute ASCR, 25068 \v{R}e\v{z}, 
Czech Republic}
\vspace{-3mm}
\from{$^{f}$\,Departamento de F\'isica, CFM, Universidade Federal 
de Santa Catarina,\\ C.P. 476, CEP 88.040-900, Florian\'opolis, 
SC, Brazil}

\begin{abstract}
A model independent L\'evy expansion method is introduced to describe nearly L\'evy shaped squared moduli  of Fourier transforms. We apply this method to precisely characterize the most recent elastic scattering data of proton-proton collisions at $\sqrt{s} = 13$ TeV. The results reveal a  substructure of protons, which is found to be significantly larger and darker in high energy proton-proton collisions at the TeV scale, as compared to a rather faint and apparently overlooked substructure, that we also identify in elastic proton-proton scattering at the ISR energy range of $\sqrt{s} =  23 - 62$ GeV.
\end{abstract}

\vspace*{6pt}

\noindent


\vspace*{12pt}

\label{sec:intro}
\section*{Introduction}

%{\it Introduction --} 
L\'evy stable distributions are ubiquitous in Nature. They correspond to the applicability of generalized central limit theorems, when the final distribution is a convolution of several apparently random elementary processes~\cite{Uchaikin:1999uz,Tsallis:1995zz}. In  high energy particle physics, squared modulus of a Fourier-transformed source distribution may have an approximately L\'evy (or, stretched exponential) shape for example in the analysis of the two-particle Bose-Einstein correlation functions~\cite{Csorgo:2003uv}, as well as in the case of the differential cross-section of elastic scattering at low four-momentum transfers~\cite{Csorgo:2018uyp}.

Recently, the TOTEM collaboration presented the data on elastic proton-proton scattering~\cite{Antchev:2018edk} at the Large Hadron Collider (LHC) energy of $\sqrt{s}= 13 $ TeV, the currently largest collision energy ever achieved in a laboratory experiment. The precision and the range of this data set provides a challenge for a statistically acceptable, quantitative description, which is a precondition for a reliable imaging process that aims to determine also the statistical significance of the analysis results.

Here we present a precise imaging of this TOTEM data set with the help of a novel and model-independent L\'evy expansion method, that  has its roots in precision analysis of Bose-Einstein correlation functions in high energy particle and heavy-ion physics~\cite{Csorgo:1999wx,DeKock:2012gp,Novak:2016cyc}.  The results reveal not only the structure (the impact parameter dependent shadow or inelastic profile function) of the protons on the femtometer scale, but for the first time the shadow profiles of a substructure inside the protons is also reconstructed and its energy evolution is investigated from the top TeV LHC collision energy of $\sqrt{s} = $ 13 TeV down to the lowest ISR energy of  23.5 GeV.

\label{sec:formalism}
\section*{Formalism}

%{\it Formalism --} 
The differential cross-section of elastic scattering is proportional to the modulus square of the elastic scattering amplitude. The TOTEM Collaboration established~\cite{Antchev:2018edk,Antchev:2015zza}, that this differential cross-section deviates significantly  from a simple exponential shape, $d\sigma/dt = A \exp(- B|t|)$ 
at low values of the four-momentum transfer $t= (p_1 - p_3)^2 $. We introduce the exponent $\alpha$ to characterize this deviation by a single parameter, so we assume that to leading order, $d\sigma/dt = A \exp\left[- (R^2 |t|)^{\alpha}\right]$. This form is frequently called as the Fourier-transform of a symmetric L\'evy-stable source distribution, sometimes also called as a stretched exponential distribution. We have recently shown that this approximation with fixed $\alpha = 0.9$ gives a statistically acceptable description of proton-proton elastic scattering at low-$t$ from $\sqrt{s} = $ 23.5 GeV to 13 TeV~\cite{Csorgo:2018uyp}.

At larger absolute values of the four-momentum transfer $t$, an interference or dip-bump structure is also apparent in these differential cross-sections, that indicates a rather significant deviation from the symmetric L\'evy source distributions and from the approximate applicability of generalized central limit theorems. 

To quantify the deviations from a symmetric L\'evy source distribution, let us utilize an orthonormal series expansion, with complex coefficients $c_i$, where the L\'evy polynomials are introduced with the help of a Gram-Schmidt orthonormalization as follows \cite{Csorgo:2018uyp}:
\begin{eqnarray} %
  \frac{d\sigma}{dt}  =    A \, w(z|\alpha) 
   \Big| 1 +
   \sum_{j = 1}^\infty c_j l_j (z|\alpha) \Big|^2\,, \quad w(z|\alpha)  =  e^{-z^\alpha}, \quad
   z  =  |t| R^2 \geq 0\,,
   \label{e:levy-dsigmadt} 
\end{eqnarray}
where $c_j  =  a_j + i \, b_j$, $w(z|\alpha)$ stands for the  L\'evy stable weight function, $R$ is the L\'evy scale parameter 
and $l_j(z|\alpha)$ stands for the normalized L\'evy polynomial of order $j$.

The $n$-th moment of $w(z|\alpha)$, $\mu_{n}^{\alpha}$ is written in terms of the Euler's gamma function as
\begin{equation}
    \mu_{n}^{\alpha} = \int_0^\infty dz\;z^{n} e^{-z^\alpha} = \frac{1}{\alpha}\,\Gamma\left( \frac{n+1}{\alpha}\right) \,, \quad
    \Gamma(x) = \int_0^\infty dz\;z^{x-1}e^{-z}
\end{equation}
The non-trivial Gram-determinants $D_j \equiv D_j(\alpha)$ (with $D_0(\alpha) =  1$) are
\begin{eqnarray} %
D_1(\alpha)  =  \mu_{0}^{\alpha} \,, \quad D_2(\alpha)  = 
     \det\left(\begin{array}{c@{\hspace*{8pt}}c}
     \mu_{0}^{\alpha} & \mu_{1}^{\alpha}  \\ 
     \mu_{1}^{\alpha} & \mu_{2}^{\alpha} 
     \end{array} \right)\,, \quad
D_3(\alpha) = 
     \det\left(\begin{array}{c@{\hspace*{8pt}}c@{\hspace*{8pt}}c}
     \mu_{0}^{\alpha} & \mu_{1}^{\alpha} & \mu_{2}^{\alpha} \\ 
     \mu_{1}^{\alpha} & \mu_{2}^{\alpha} & \mu_{3}^{\alpha} \\ 
     \mu_{2}^{\alpha} & \mu_{3}^{\alpha} & \mu_{4}^{\alpha} 
       \end{array} \right), \, ...  \, \mbox{\rm etc} \,. \label{e:gram}
\end{eqnarray}

The normalized L\'evy polynomials $l_j(z|\alpha)$ are introduced in terms of the Gram-determinants and the unnormalized L\'evy polynomials~\cite{Csorgo:2018uyp} as follows:
\begin{eqnarray}
        l_j(z\, |\, \alpha) & = & D^{-\frac{1}{2}}_{j} D^{-\frac{1}{2}}_{j+1} 
        L_j(z\, |\, \alpha) \,, \quad \mbox{\rm for}\quad j\ge 0 \, .
    \label{e:lj}
\end{eqnarray}
In this equation the  unnormalized L\'evy polynomials are denoted as $L_i(z\,|\,\alpha)$ (with $L_0(z\,|\,\alpha) =  1$). These polynomials 
were introduced in Ref.~\cite{Novak:2016cyc} and are defined as follows:
\begin{eqnarray}
 L_1(z\,|\,\alpha)  = 
       \det\left(\begin{array}{c@{\hspace*{8pt}}c}
     \mu_{0}^{\alpha} & \mu_{1}^{\alpha}  \\ 
     1 & z \end{array} \right) \,, \quad
 L_2(z\,|\,\alpha) = 
       \det\left(\begin{array}{c@{\hspace*{8pt}}c@{\hspace*{8pt}}c}
     \mu_{0}^{\alpha} & \mu_{1}^{\alpha} & \mu_{2}^{\alpha} \\ 
     \mu_{1}^{\alpha} & \mu_{2}^{\alpha} & \mu_{3}^{\alpha}  \\ 
     1 & z & z^2 \end{array} \right), \, ... \,
          \mbox{\rm etc} \, . 
\end{eqnarray}

\label{sec:observables}
\section*{Observables}

%{\it Observables --} 
The elastic differential cross-section is related to the modulus squared Fourier transformed elastic scattering amplitude $T_{el}(\Delta)$ as
\begin{eqnarray}
\frac{d\sigma}{dt} = \frac{1}{4\pi}|T_{el}(\Delta)|^2 \,, \qquad \Delta=\sqrt{|t|}\, . 
\label{e:dsigmadt-Tel}
\end{eqnarray}
We may assume, that the elastic scattering amplitude is predominantly imaginary at high energies. In this case,  the zeroth order normalization coefficient can be absorbed to the overall normalization coefficient $A$, and the following  L\'evy expansion is obtained for the complex-valued elastic scattering amplitude $T_{el}(\Delta)$:
\begin{eqnarray}
T_{el}(\Delta) & = & 
    i\sqrt{4\pi A}\, e^{- z^{\alpha}/2}  
                \!  \left(\!
                1+\sum_{i = 1}^\infty c_i l_i (z|\alpha) \! \right)_{z = \Delta^2 R^2} \!\!\!\!\! . 
        \label{e:Tel}
\end{eqnarray}
The total and elastic cross-sections are given as
\begin{eqnarray}
\sigma_{\rm tot}  \equiv  2\,{\rm Im}\, T_{el}(0)  = 
2\,\sqrt{4\pi A}\,
        \left(1 + \sum_{i = 1}^\infty a_i l_i (0|\alpha) \right) \!\! , & \null &\null
    \label{e:sigmatot} \\
    \sigma_{\rm el}  \equiv  \!\! \int_0^\infty \!\!\! d|t| \frac{\mbox{\rm d}\sigma}{\mbox{\rm d} t}  = 
    \frac{A}{R^2}\left[\frac{1}{\alpha}\Gamma\left(\frac{1}{\alpha}\right) + \sum_{i = 1}^\infty (a_i^2 + b_i^2) \right] \!\! . &  \null & \null
    \label{e:sigmael}
\end{eqnarray}

\label{sec:imaging}
\section*{Imaging on the femtometer scale}

%{\it Imaging on the femtometer scale --} 
The angular distribution of elastic scattering provides pictures of the elastically scattered particles in impact parameter 
or $\mathbf b$ space. For spin independent processes,
\begin{equation}
        t_{el}(b)=
        \frac{1}{2\pi}
        \int J_0(\Delta\,b)\,T_{el}(\Delta)\,\Delta\, d\Delta \, ,
                \label{tel-b}
\end{equation}       
is the impact parameter dependent elastic scattering amplitude,
where $ \Delta\equiv|\mathbf{\Delta}|$,  $b\equiv|\mathbf{b}|\, $ 
and $J_0(x)$ is the zeroth-order Bessel function of the first kind.
This $t_{el}(b)$ can also be represented in an eikonal form as
\begin{eqnarray}
        t_{el}(b)=i\left[ 1 - e^{-\Omega(b)} \right] \,, \qquad  P(b) = 1-\left|e^{-\Omega(b)}\right|^2 \,.
        \label{e:tel-eikonal}
\end{eqnarray}
where $\Omega(b) $ is the so-called opacity function (known also as the eikonal), which is in general complex, and the proton image 
is given by the shadow (or inelasticity) profile function $P(b)$. This way,  the rather fundamental results of multiple diffraction theory~\cite{Glauber:1984su} when powered with the help of the newly found L\'evy expansion of the elastic scattering amplitude, together make it possible now to perform a high precision optical imaging of elastic hadron-hadron scattering and obtain pictures of the (sub)structure(s) of the protons on the femtometer scale.
%%%%%%%%%%%%%%%%%%%%%%%%%%%%%%%%%%%%%%%%%%
\begin{figure}[t]
\begin{center}
        \includegraphics[scale=0.36]{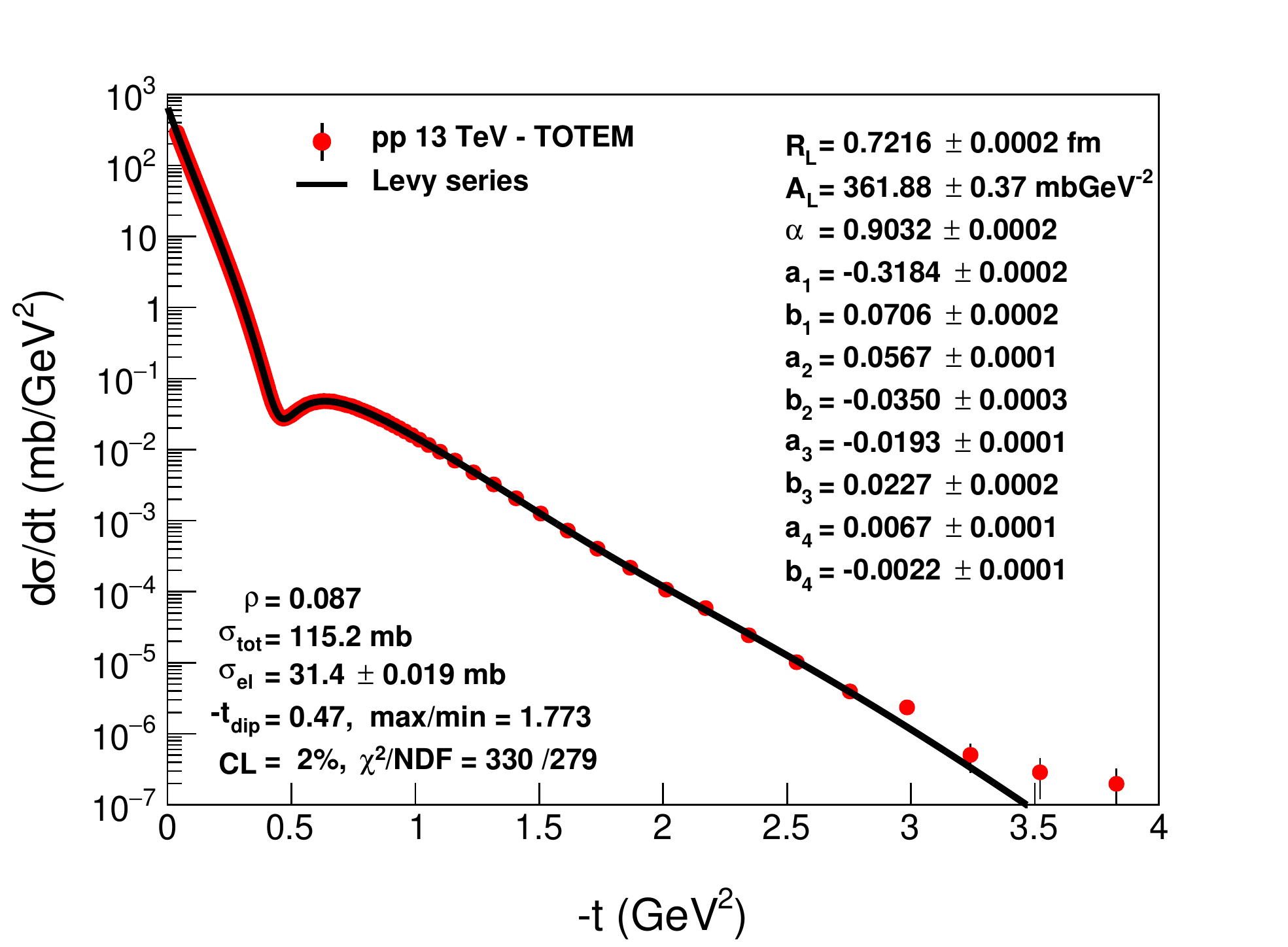}
        \includegraphics[scale=0.30]{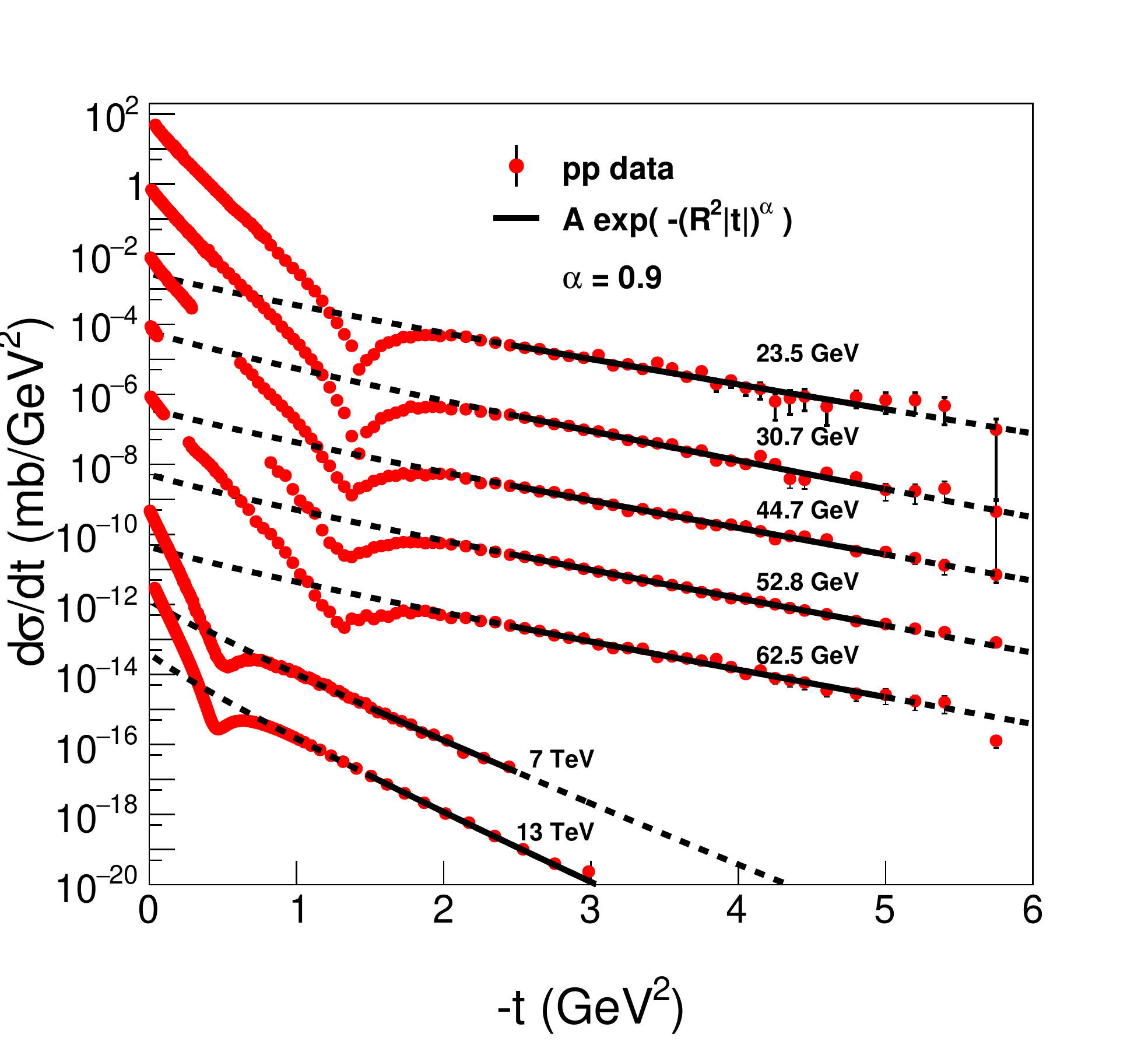}
        \caption{{\bf Left panel:} Eq.~(\ref{e:levy-dsigmadt}) fitted to TOTEM  $d\sigma/dt$ data at $\sqrt{s} = 13$ TeV in elastic proton-proton collisions. 
        {\bf Right panel:} Eq.~(\ref{e:levy-dsigmadt}) with $(a_i,b_i) \equiv (0,0)$ for $i \ge 1$ describes the nearly exponential behaviour of $d\sigma/dt$ of elastic 
        proton-proton collisions, with $\alpha = 0.9$ fixed, in the secondary cone after the dip and the bump, corresponding to $2.5 \le -t \le 5$ GeV$^2$ at the ISR 
        and  $1.5 \le -t \le 3$ GeV$^2$ at LHC energies, with parameters of Tables~\ref{t:sub-ISR} and \ref{t:sub-LHC}.
        }         \label{fig:levyfit-13TeV_alltails}
\end{center}
\end{figure}
%%%%%%%%%%%%%%%%%%%%%%%%%%%%%%%%%%%%%%%%%%
%%%%%%%%%%%%%%%%%%%%%%%%%%%%%%%%%%%%%%%%%%
\begin{figure}[tb]
%\begin{center}
\begin{minipage}{0.5\textwidth}
 \includegraphics[scale=0.45]{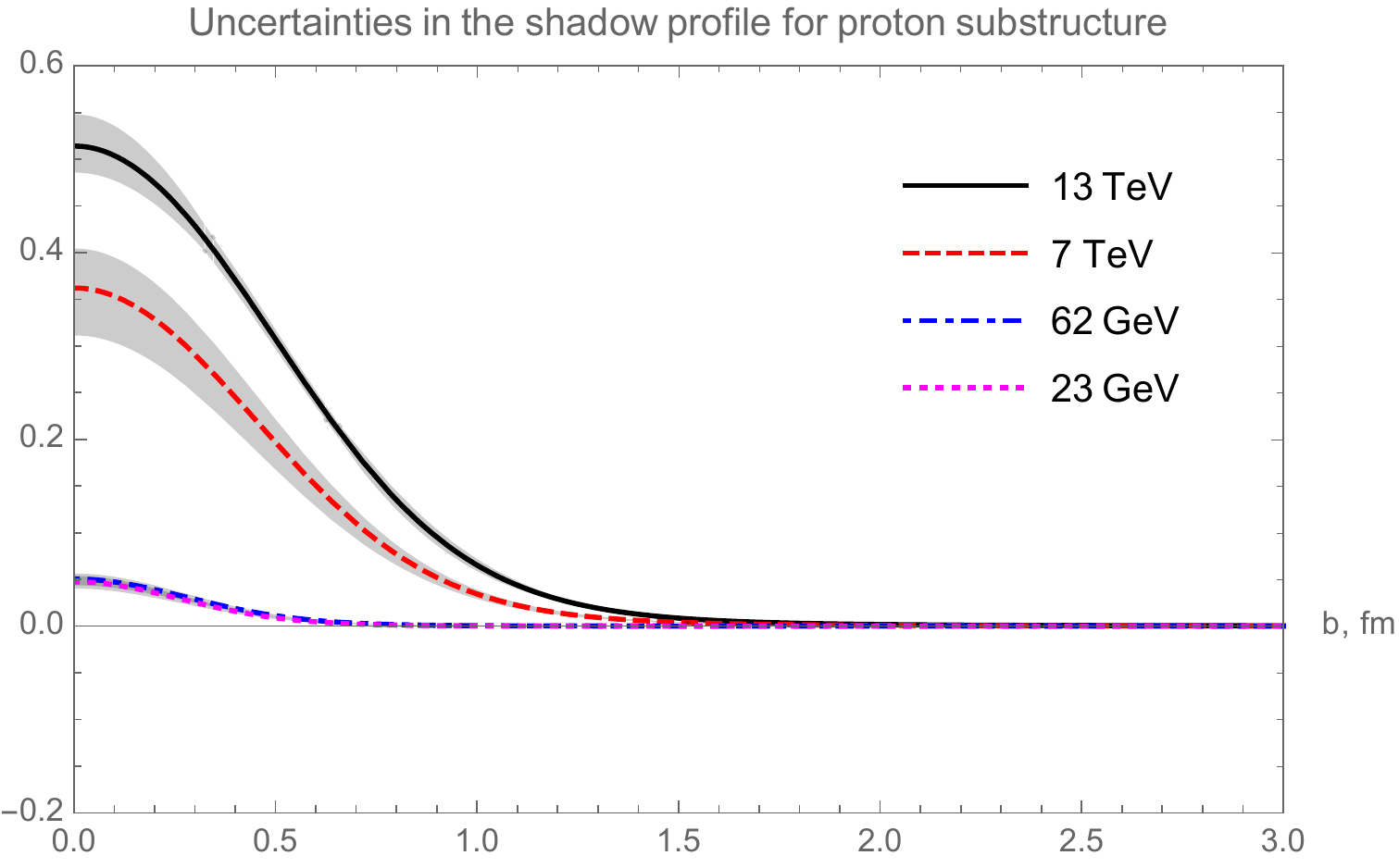}
\end{minipage}
\begin{minipage}{0.5\textwidth}
  \includegraphics[scale=0.45]{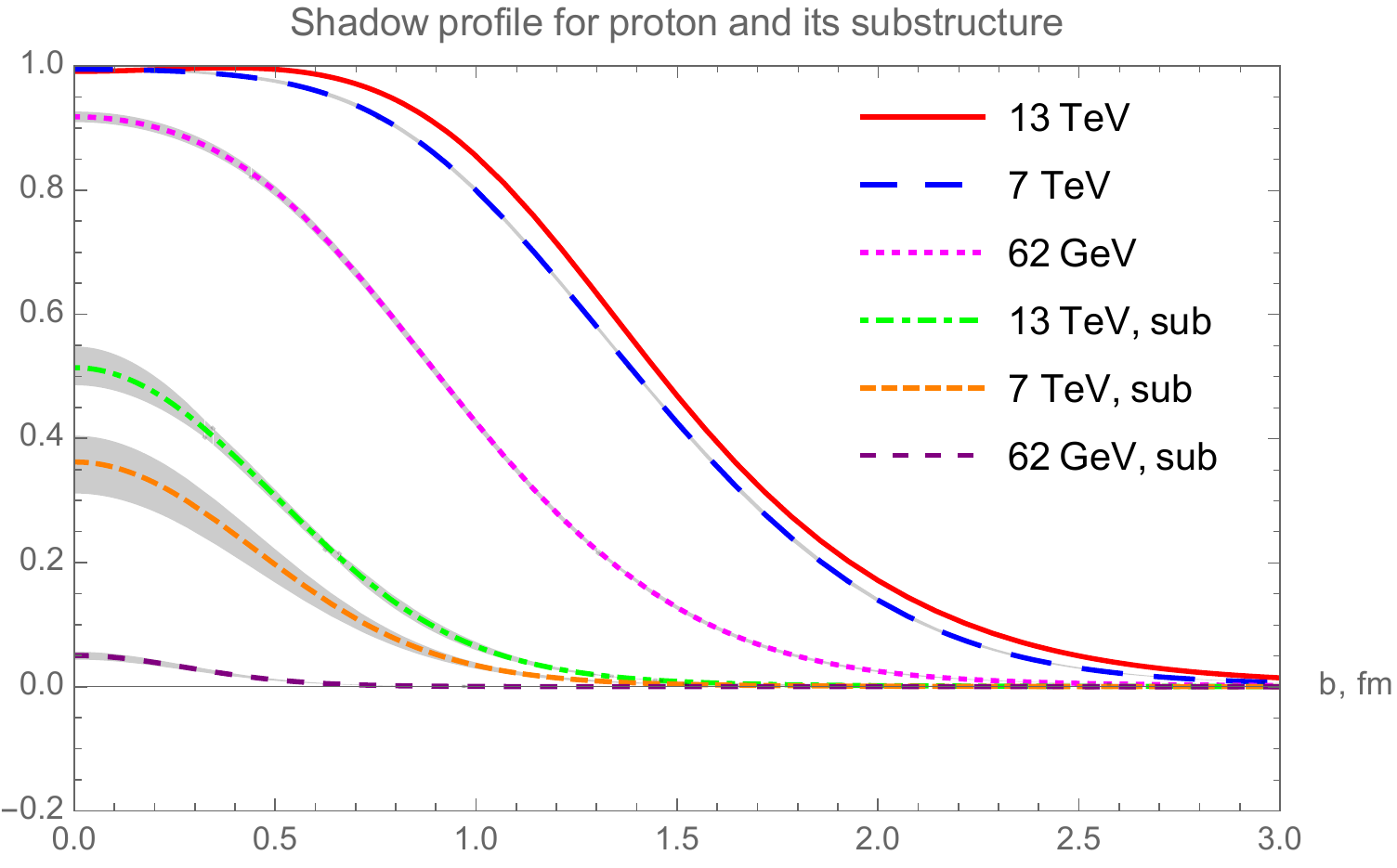} 
\end{minipage}
        \caption{
        {\bf Left panel:} Images and the imaging errors of the internal substructures inside 
        the protons at various collision energies, including error bars.
        {\bf Right panel:} Images of protons and their substructures at various collision energies, including error bars.
        The shadow profile of the protons is growing with increasing collision energies.
        }
        \label{fig:shadow-all}
\end{figure}
%%%%%%%%%%%%%%%%%%%%%%%%%%%%%%%%%%%%%%%%%%
%%%%%%%%%%%%%%%%%%%%%%%%%%%%%%%%%%%%%%%%%%
\begin{table*}[t]
    \centering
\scalebox{.8}{    \begin{tabular}{c|c|c|c|c|c} %
       $\sqrt{s}$ (GeV)   & 23.5 & 30.7 & 44.7 & 52.8 & 62.5 
       \\ \hline
         A (mb GeV$^2$) & 0.003 $\pm$ 0.001 &
           0.006 $\pm$ 0.001 & 
               0.004 $\pm$ 0.001 & 
                    0.005 $\pm$ $\pm$ 0.001& 
                        0.004 $\pm$ 0.001 \\
         R (fm) & 0.30 $\pm$ 0.01 & 
            0.32 $\pm$ 0.01& 
                0.31 $\pm$ 0.01 & 
                    0.32 $\pm$ 0.01& 
                        0.31 $\pm$ 0.01  \\
         $\chi^2/{\rm NDF}$ & 22.7 / 21 & 
            16.3 / 21 & 
                19 / 21 & 
                    13.9 / 21 & 
                        12.2 / 21 \\
         CL [\%] &  36 & 
            75 & 
                59 & 
                   87 & 
                        94  \\ \hline
         $\sigma^{\mbox{\rm sub}}_{\mbox{\rm tot}}$ (mb)  & 0.24 $\pm$ 0.03 & 
                0.35 $\pm$ 0.03    & 
                       0.28 $\pm$ 0.02 & 
                            0.32 $\pm$ 0.01 & 
                                0.30 $\pm$ 0.03  \\ 
 $\sigma^{\mbox{\rm sub}}_{\mbox{\rm el}}$ ($\mu$b)  & 
        1.3 $\pm$ 0.3 & 
                2.4 $\pm$ 0.4    & 
                       1.7 $\pm$ 0.2 & 
                            2.1 $\pm$ 0.2 & 
                                1.9 $\pm$ 0.3  %
    \end{tabular} }
    \caption{Substructure parameters from L\'evy fits, $d\sigma/dt = A \exp(-(|t|R^2)^{\alpha})$ at various ISR collision energies, for $\alpha = 0.9$.}
    \label{t:sub-ISR}
\end{table*}
%%%%%%%%%%%%%%%%%%%%%%%%%%%%%%%%%%%%%%%%%%
%%%%%%%%%%%%%%%%%%%%%%%%%%%%%%%%%%%%%%%%%%
\begin{table}[t]
    \centering
\scalebox{.8}{    \begin{tabular}{c|c|c} %
       $\sqrt{s}$ (TeV)   & 7 & 13   
       \\ \hline
         A (mb GeV$^2$) & 1.4 $\pm $ 0.4 &
           4.0 $\pm$ 0.3 \\
         R (fm) & 0.49 $\pm$ 0.01 & 
            0.51 $\pm$ 0.01 \\
         $\chi^2/{\rm NDF}$ & 4.9 / 10 & 
            16.9 / 8 \\
         CL [\%] &  90 & 
           3  \\ \hline
         $\sigma^{\mbox{\rm sub}}_{\mbox{\rm tot}}$ (mb)  & 5.2  $\pm$ 0.7  & 
                8.9 $\pm$ 0.30    \\ 
         $\sigma^{\mbox{\rm sub}}_{\mbox{\rm el}}$ (mb)  & 0.24  $\pm$ 0.06  & 
                0.63 $\pm$ 0.04   
    \end{tabular} }
    \caption{Substructure parameters from L\'evy fits, $d\sigma/dt = A \exp(-(|t|R^2)^{\alpha})$ at LHC collision energies, for $\alpha = 0.9$.}
    \label{t:sub-LHC}
\end{table}
%%%%%%%%%%%%%%%%%%%%%%%%%%%%%%%%%%%%%%%%%%

\label{sec:results}
\section*{Results}

%{\it Results --} 
The model-independent L\'evy expansion method is shown to describe  the recently presented TOTEM data~\cite{Antchev:2018edk} of elastic proton-proton scattering at $\sqrt{s} = $ 13 TeV in Fig.~\ref{fig:levyfit-13TeV_alltails} (left panel). These data span 10 orders of magnitude on the vertical scale, indicate a nearly exponential cone region at low values of the four-momentum transfer $t  $, that is followed by a dip and bump structure and a secondary tail. The L\'evy expansion has a confidence level CL = 2 \%. This indicates  that the L\'evy expansion represents these data in an acceptable manner. As far as we know,  the precision of this data set is unprecedented in the field of elastic scattering in high energy physics and its description on a quantitative,  statistically acceptable manner remains a challenge at present for all other, model dependent attempts. At $\sqrt{s} = 13$ TeV, the maximal measured four-momentum transfer values are $-t_{max} \approx 4 $ GeV$^2$. This allows for a minimal spatial resolution of  $\hbar/\sqrt{-t_{max}} \approx 0.1 $ fm, and makes it possible to resolve some of the substructures inside the colliding protons at LHC energies. The model independent L\'evy expansion technique is also well suited to describe earlier measurements of elastic proton-proton scattering with $\sqrt{s}$ ranging from  23.5 GeV to 7 TeV, as detailed in Appendix A of Ref.~\cite{Csorgo:2018uyp}.

Each of the differential cross sections, summarized in Fig.~\ref{fig:levyfit-13TeV_alltails} (right panel),
lack secondary dip and bump structure. As far as we know, the significance of this observation was  noted first in Ref.~\cite{Dremin:2018uwt}. As we  detailed in Ref.~\cite{Csorgo:2018uyp}, the data at $-t $ larger than the dip and bump region are described in a statistically acceptable manner with a zeroth order L\'evy fit, corresponding to $d\sigma/dt = A \exp(- (|t|R^2)^{\alpha}$, with $\alpha = 0.9$ fixed value. This value of $\alpha$ indicates a small but systematic deviation from the exponential shape. The results, summarized in Fig.~\ref{fig:levyfit-13TeV_alltails} (right panel), reveal another striking and unexpected detail: at the ISR energy range of 23.5 $\le \sqrt{ s}\le$ 62.5 GeV the fits with the zeroth order L\'evy expansion yield nearly parallel lines, indicating that neither the L\'evy exponent $\alpha$ nor the L\'evy scale $R$ of this tail region evolve significantly with changing the collision energy in this lower energy range.

A corresponding  substructure is reconstructed inside the proton with the help of Eq.~(\ref{e:tel-eikonal}), as indicated in the left panel of  Fig.~\ref{fig:shadow-all}. At ISR energies, this  substructure is very faint, with maximum value of $P(b=0)$ of about 0.05. Apparently, there are no statistically significant differences among the  faint shadows of these proton substructures in this ISR energy region of $\sqrt{s} = 23.5$ to 62.5 GeV. A significantly different, larger and darker substructure is however found in elastic proton-proton collisions at 7 TeV. Increasing the center of mass energy from 7 to 13 TeV, the same kind of larger and darker substructure is observed again, with a size that does not grow significantly from 7 TeV to 13 TeV, but has a significantly darker shadow, as shown in
the left panel of Fig.~\ref{fig:shadow-all}. The zeroth order L\'evy fit parameters that characterize the secondary cone region and the corresponding substructures at various collision energies are summarized in Tables ~\ref{t:sub-ISR} and \ref{t:sub-LHC}. The calculated contributions of these substructures to the total and to the elastic cross-section, $\sigma_{tot}^{sub}$ and $\sigma_{el}^{sub}$ are also shown, as  obtained from Eqs.~(\ref{e:sigmatot}) and (\ref{e:sigmael}) with $(a_i,b_i) = (0,0)$.

The right panel of Figure~\ref{fig:shadow-all} indicates the shadow profiles of the protons at $\sqrt{s} = $ 13 TeV, 7 TeV and 62 GeV, and compares them with the shadow profiles of the substructures of the protons at the same energies. The shadow profiles of these substructures are significantly smaller and fainter  than the shadow of the protons at each considered energy. With increasing collision energy, the size and the darkness of the proton increase. In the LHC energy range, a nearly constant maximally dark region develops in the central region of the proton shadow, that grows with increasing energy. This structure is qualitatively different from the nearly Gaussian proton structure apparent in the collisions at lower energies. Surprisingly, a substructure is also resolved in the ISR energy range: it is found to be rather faint and its size is a constant within the errors of resolution. At LHC energies, the imaged substructure is significantly larger and darker as compared to that at  ISR energies.

\label{sec:summary}
\section*{Summary}

%{\it Summary --} 
To summarise, we have imaged the overall space-time structure of elastic proton-proton collisions at the largest currently available collision energy of $\sqrt{s}= 13 $ TeV and CERN LHC, using the recent TOTEM data~\cite{Antchev:2018edk}. Uniting  Glauber's multiple diffraction theory with the new orthonormal L\'evy series expansion of the elastic scattering amplitude, we have achieved an unprecedented precision in describing the differential cross section of elastic proton-proton scattering, over 10 orders of magnitude. Surprisingly, we have identified and quantified a substructure of the protons, that contributes to the total cross-section with about 0.3 mb in the energy range of $\sqrt{s}= 23.5 -62.5$ ISR energy range and about 5.2 and 8.9 mb at $\sqrt{s} = 7$ and 13 TeV LHC energies. 

The imaging method presented here can only report on the observation of these substructures but, being model independent, it cannot be overused to determine the detailed physical meaning of the observed signatures. Nevertheless, let us mention that these are strikingly similar in magnitude to the size of the dressed quarks and diquarks that have been determined from a model dependent analysis of proton-proton elastic scattering at the ISR and LHC energies~\cite{Bialas:2006qf,Nemes:2012cp,Csorgo:2013bwa,CsorgO:2013kua,Csorgo:2013bwa,Nemes:2015iia}. The similarity of the model independent L\'evy imaging results to the results of this Bialas-Bzdak quark-diquark picture of elastic proton-proton scattering may also provide a quantitative support to the quark-diquark picture of baryons, that were recently obtained by Brodsky and collaborators using AdS/QCD strong coupling calculations~\cite{Brodsky:2017qno}. While our model independent imaging results are consistent with this QCD based interpretation, they are open for other possible quantitative models and descriptions to explain the origin of the observed signatures.

{\it Acknowledgments}
T. Cs. thanks to R.J. Glauber for inspiring discussions and for proposing a series expansion to describe elastic scattering. We acknowledge inspiring  discussions with 
S. Giani, W. Guryn, G. Gustaf\-son, L. L\"onnblad, K. \"Osterberg and M. \v{S}umbera and thank the TOTEM Collaboration for making their preliminary data set available for us. 
R. P. is partially supported by the Swedish Research Council grants No. 621-2013-4287 and 2016-05996, by an ERC H2020 grant No 668679, as well as by the Ministry of Education, Youth and Sports of the Czech Republic project LT17018. T. Cs. and A. S. were partially supported by the NKIFH grants No. FK-123842 and FK-123959, and by the  EFOP 3.6.1-16-2016-00001 grant (Hungary) and the NKM-92/2017 projects of the Hungarian  and the Ukrainian Academy of Sciences. Our collaboration was supported by THOR, the EU COST Action CA15213. 

\bibliographystyle{pepan}
\bibliography{biblio}

\end{document}